\documentclass[10pt,letterpaper]{article}
\usepackage{opex3}

\usepackage{amsmath}
\usepackage{amssymb}
\usepackage{graphicx}
\usepackage{color}
\usepackage{bm}
\usepackage{latexsym}
\usepackage{empheq}
\usepackage{braket}
\usepackage{appendix}

\newcommand{\ch}{\ensuremath{\chi^{(3)}}} 
\newcommand{\vE}{\ensuremath{\mathbf{E}}}

\renewcommand{\eqref}[1]{Eq.~\ref{eq:#1}}
\newcommand{\eqreftwo}[2]{Eqs.~\ref{eq:#1}--\ref{eq:#2}}

\newcommand{\figref}[1]{Fig.~\ref{fig:#1}}
\newcommand{\Figref}[1]{Figure~\ref{fig:#1}}

\begin{document}

\title{Design of diamond microcavities for single photon frequency down-conversion}
\author{Z. Lin, S. G. Johnson, A. W. Rodriguez and M. Loncar}
\address{School of Engineering and Applied Sciences, Harvard University, Cambridge MA, 02138}

\email{zlin@seas.harvard.edu} 


\begin{abstract}
We propose monolithic diamond cavities that can be used to convert color-center Fock-state single photons from emission wavelengths to telecommunication bands. We present a detailed theoretical description of the conversion process, analyzing important practical concerns such as nonlinear phase shifts and frequency mismatch. Our analysis predicts sustainable power requirements ($ \lesssim 1~\mathrm{W}$) for a chipscale nonlinear device with high conversion efficiencies.
\end{abstract}


\bibliographystyle{unsrt}
\bibliography{photon}

\begin{thebibliography}{10}

\bibitem{Michler00}
P.~Michler, A.~Kiraz, C.~Becher, W.~V. Schoenfeld, P.~M. Petroff, L.~Zhang,
  E.~Hu, and A.~Imamoglu.
\newblock A quantum dot single-photon turnstile device.
\newblock {\em Science}, 290:2282--2285, 2000.

\bibitem{Pelton02}
M.~Pelton, C.~Santori, J.~Vu{\v{c}}kovi{\'{c}}, B.~Zhang, G.~S. Solomon,
  J.~Plant, and Y.~Yamamoto.
\newblock Efficient source of single photons: a single quantum dot in a
  micropost microcavity.
\newblock {\em Phys. Rev. Lett.}, 89(23):233602, 2002.

\bibitem{Duan04}
L.~M. Duan and H.~J. Kimble.
\newblock Scalable photonic quantum computation through cavity-assisted
  interactions.
\newblock {\em Phys. Rev. Lett.}, 94(12):127902, 2004.

\bibitem{Cirac97}
J.~I. Cirac, P.~Zoller, H.~J. Kimble, and H.~Mabuchi.
\newblock Quantum state transfer and entanglement distribution among distant
  nodes in a quantum network.
\newblock {\em Phys. Rev. Lett.}, 78:3221--3224, 1997.

\bibitem{Kuhn02}
A.~Kuhn, M.~Hennrich, and G.~Rempe.
\newblock Deterministic single-photon source for distributed quantum
  networking.
\newblock {\em Phys. Rev. Lett.}, 89(6):067901, 2002.

\bibitem{Childress06}
L.~Childress, J.~M. Taylor, A.~S. Sorensen, and M.~D. Lukin.
\newblock Fault-tolerant quantum communication based on solid-state photon
  emitters.
\newblock {\em Phys. Rev. Lett.}, 96:070504, 2006.

\bibitem{Kimble08}
H.~J. Kimble.
\newblock The quantum internet.
\newblock {\em Nature}, 453:1023--1030, 2008.

\bibitem{Kurtsiefer00}
C.~Kurtsiefer, S.~Mayer, P.~Zarda, and H.~Weinfurter.
\newblock Stable solid-state source of single photons.
\newblock {\em Phys. Rev. Lett.}, 85:290--293, 2000.

\bibitem{Beveratos02}
A.~Beveratos, R.~Brouri, T.~Gacoin, A.~Villing, J.~Poizat, and P.~Grangier.
\newblock Single photon quantum cryptography.
\newblock {\em Phys. Rev. Lett.}, 89:187901, 2002.

\bibitem{Babinec10}
T.~M. Babinec, B.~J. Hausmann, M.~Khan, Y.~Zhang, J.~R. Maze, P.~R. Hemmer, and
  M.~Loncar.
\newblock A diamond nanowire single-photon source.
\newblock {\em Nat. Nano.}, 5:195--199, 2010.

\bibitem{Jelezko04}
F.~Jelezko, T.~Gaebel, I.~Popa, A.~Gruber, and J.~Wrachtrup.
\newblock Observation of coherent oscillations in a single electron spin.
\newblock {\em Phys. Rev. Lett.}, 92:076401, Feb 2004.

\bibitem{Prawer08}
Steven Prawer and Andrew~D. Greentree.
\newblock Diamond for quantum computing.
\newblock {\em Science}, 320(5883):1601--1602, 2008.

\bibitem{Dutt07}
M.~V.~Gurudev Dutt, L.~Childress, L.~Jiang, E.~Togan, J.~Maze, F.~Jelezko,
  A.~S. Zibrov, P.~R. Hemmer, and M.~D. Lukin.
\newblock Quantum register based on individual electronic and nuclear spin
  qubits in diamond.
\newblock {\em Science}, 316(5829):1312--1316, 2007.

\bibitem{Neumann08}
P.~Neumann, N.~Mizuochi, F.~Rempp, P.~Hemmer, H.~Watanabe, S.~Yamasaki,
  V.~Jacques, T.~Gaebel, F.~Jelezko, and J.~Wrachtrup.
\newblock Multipartite entanglement among single spins in diamond.
\newblock {\em Science}, 320(5881):1326--1329, 2008.

\bibitem{McCutcheon08}
M.~W. McCutcheon and M.~Loncar.
\newblock Design of a silicon nitride photonic crystal nanocavity with a
  quality factor of one million for coupling to a diamond nanocrystal.
\newblock {\em Opt. Express}, 16(23):19136, 2008.

\bibitem{Prawer09}
Igor Aharonovich, Chunyuan Zhou, Alastair Stacey, Julius Orwa, Stefania
  Castelletto, David Simpson, Andrew~D. Greentree, Fran\ifmmode
  \mbox{\c{c}}\else~\c{c}\fi{}ois Treussart, Jean-Francois Roch, and Steven
  Prawer.
\newblock Enhanced single-photon emission in the near infrared from a diamond
  color center.
\newblock {\em Phys. Rev. B}, 79:235316, Jun 2009.

\bibitem{Prawer11}
I.~Aharonovich, S.~Castelletto, B.~C. Johnson, J.~C. McCallum, and S.~Prawer.
\newblock Engineering chromium-related single photon emitters in single crystal
  diamonds.
\newblock {\em New Journal of Physics}, 13(045015), 2011.

\bibitem{Jelezko13}
L.~J. {Rogers}, K.~D. {Jahnke}, L.~{Marseglia}, C.~{M{\"u}ller}, B.~{Naydenov},
  H.~{Schauffert}, C.~{Kranz}, T.~{Teraji}, J.~{Isoya}, L.~P. {McGuinness}, and
  F.~{Jelezko}.
\newblock {Creation of multiple identical single photon emitters in diamond}.
\newblock {\em ArXiv e-prints}, Oct 2013.

\bibitem{Benson14}
Matthias Leifgen, Tim Schröder, Friedemann Gädeke, Robert Riemann, Valentin
  Métillon, Elke Neu, Christian Hepp, Carsten Arend, Christoph Becher,
  Kristian Lauritsen, and Oliver Benson.
\newblock Evaluation of nitrogen- and silicon-vacancy defect centres as single
  photon sources in quantum key distribution.
\newblock {\em New Journal of Physics}, 16(2):023021, 2014.

\bibitem{Becher12}
Elke Neu, Mario Agio, and Christoph Becher.
\newblock Photophysics of single silicon vacancy centers in diamond:
  implications for single photon emission.
\newblock {\em Opt. Express}, 20(18):19956--19971, Aug 2012.

\bibitem{Becher14}
Christian Hepp, Tina M\"uller, Victor Waselowski, Jonas~N. Becker, Benjamin
  Pingault, Hadwig Sternschulte, Doris Steinm\"uller-Nethl, Adam Gali,
  Jeronimo~R. Maze, Mete Atat\"ure, and Christoph Becher.
\newblock Electronic structure of the silicon vacancy color center in diamond.
\newblock {\em Phys. Rev. Lett.}, 112:036405, Jan 2014.

\bibitem{Sipahigil14}
A.~Sipahigil, K. D. Jahnke, L. J. Rogers, T.~Teraji, J.~Isoya, A. S.
  Zibrov, F.~Jelezko, and M. D. Lukin.
\newblock Indistinguishable photons from separated silicon-vacancy centers in
  diamond.
\newblock {\em Phys. Rev. Lett.}, 113(113602), 2014.

\bibitem{Englund10}
D.~Englund, B.~Shields, K.~Rivoire, F.~Hatami, J.~Vu{\v{c}}kovi{\'{c}},
  H.~Park, and M.~D. Lukin.
\newblock Deterministic coupling of a single nitrogen vacancy center to a
  photonic crystal cavity.
\newblock {\em Nano Lett.}, 10:3922--3926, 2010.

\bibitem{Wolters10}
Wolters et~al.
\newblock Enhancement of the zero phonon line emission from a single nitrogen
  vacancy center in a nanodiamond via coupling to a photonic crystal cavity.
\newblock {\em Appl. Phys. Lett.}, 97:141108, 2010.

\bibitem{Sar11}
T.~van der Sar~et al.
\newblock Deterministic nanoassembly of a coupled quantum emitter-photonic
  crystal cavity system.
\newblock {\em Appl. Phys. Lett.}, 98:193103, 2011.

\bibitem{Leon12}
N.~P. de~Leon~et al.
\newblock Tailoring light-matter interaction with a nanoscale plasmon
  resonator.
\newblock {\em Phys. Rev. Lett.}, 108:226803, 2012.

\bibitem{Faraon11}
A.~Faraon, P.~E. Barclay, C.~Santori, K.~M.~C. Fu, and R.~G. Beausoleil.
\newblock Resonant enhancement of the zero-phonon emission from a color centre
  in a diamond cavity.
\newblock {\em Nat. Photonics}, 12:1578, 2012.

\bibitem{Hausmann12}
B.~M.~Hausmann et~al.
\newblock Integrated diamond networks for quantum nanophotonics.
\newblock {\em Nano Lett.}, 12:1578, 2012.

\bibitem{Faraon12}
A.~Faraon, C.~Santori, Z.~Huang, V.~M. Acosta, and R.~G. Beausoleil.
\newblock Coupling of nitrogen-vacancy centers to photonic crystal cavities in
  monocrystalline diamond.
\newblock {\em Phys. Rev. Lett.}, 109:033604, 2012.

\bibitem{Moeller12}
J.~Riedrich-Moeller et~al.
\newblock One- and two-dimensional photonic crystal microcavities in single
  crystal diamond.
\newblock {\em Nat. Nano.}, 7, 2012.

\bibitem{Burek12OM}
M.~J.~Burek et~al.
\newblock Free-standing mechanical and photonic nanostructures in
  single-crystal diamond.
\newblock {\em Nano Lett.}, 12:6084, 2012.

\bibitem{Hausmann13NL}
B.~J.~M. Hausmann, B.~J. Shields, Q.~Quan, Y.~Chu, N.~P. de~Leon, R.~Evans,
  M.~J. Burek, A.~S. Zibrov, M.~Markham, D.~J. Twitchen, H.~Park, M.~D. Lukin,
  and M.~Loncar.
\newblock Coupling of nv centers to photonic crystal nanobeams in diamond.
\newblock {\em Nano Lett.}, 13:5791 -- 5796, 2013.

\bibitem{Burek14}
Michael~J. Burek, Yiwen Chu, Madelaine~S.Z. Liddy, Parth Patel, Jake Rochman,
  Srujan Meesala, Wooyoung Hong, Qimin Quan, Mikhail~D. Lukin, and Marko
  Loncar.
\newblock High quality-factor optical nanocavities in bulk single-crystal
  diamond.
\newblock {\em Nat. Comm.}, 5:5718, 2014.

\bibitem{Hausmann14}
B.~J.~M. Hausmann, I.~Bulu, V.~Venkataraman, P.~Deotare, and M.~Loncar.
\newblock Diamond nonlinear photonics.
\newblock {\em Nature Photonics}, 8:369 -- 374, 2014.

\bibitem{McCutcheon09}
Murray~W. McCutcheon, Darrick~E. Chang, Yinan Zhang, Mikhail~D. Lukin, and
  Marko Loncar.
\newblock Broadband frequency conversion and shaping of single photons emitted
  from a nonlinear cavity.
\newblock {\em Opt. Express}, 17:22689, 2009.

\bibitem{Agha12}
I.~Agha, M.~Davanco, B.~Thurston, and K.~Srinivasan.
\newblock Low-noise chip-based frequency conversion by four-wave-mixing bragg
  scattering in sin$_x$ waveguides.
\newblock {\em Opt. Lett.}, 37:2997, 2012.

\bibitem{Huang13}
Y.~Huang, V.~Velev, and P.~Kumar.
\newblock Quantum frequency conversion in nonlinear microcavities.
\newblock {\em Opt. Lett.}, 38:2119--2121, 2013.

\bibitem{Drummond90}
P.~D. Drummond.
\newblock Electromagnetic quantization in dispersive inhomogeneous nonlinear
  dielectrics.
\newblock {\em Phys. Rev.~A}, 42:6845--6857, 1990.

\bibitem{Ramirez11}
D.~Ramirez, A.~W. Rodriguez, H.~Hashemi, J.~D. Joannopoulos, M.~Solijacic, and
  S.~G. Johnson.
\newblock Degenerate four-wave mixing in triply-resonant nonlinear kerr
  cavities.
\newblock {\em Phys. Rev.~A}, 83:033834, 2011.

\bibitem{Rodriguez07:OE}
Alejandro Rodriguez, Marin Solja{\v{c}}i{\'{c}}, J.~D. Joannopulos, and
  Steven~G. Johnson.
\newblock $\chi^{(2)}$ and $\chi^{(3)}$ harmonic generation at a critical power
  in inhomogeneous doubly resonant cavities.
\newblock {\em Opt. Express}, 15(12):7303--7318, 2007.

\bibitem{Johnson01:cavities}
Steven~G. Johnson, Attila Mekis, Shanhui Fan, and J.~D. Joannopoulos.
\newblock Molding the flow of light.
\newblock {\em Computing Sci. Eng.}, 3(6):38--47, 2001.

\bibitem{Hansryd02}
Jonas Hansryd, Peter~A. Andrekson, Mathias Westlund, Jie Li, and Per-Olof
  Hedekvist.
\newblock Fiber-based optical parametric amplifiers and their applications.
\newblock {\em IEEE J. Sel. Topics in Quantum Electronics}, 8(3), 2002.

\bibitem{Boyd92}
Robert~W. Boyd.
\newblock {\em Nonlinear Optics}.
\newblock Academic Press, California, 1992.

\bibitem{Neu11}
Elke Neu, David Steinmetz, Janine Riedrich-Moller, Stefan Gsell, Martin
  Fischer, Matthias Schreck, and Christoph Becher.
\newblock Single photon emission from silicon-vacancy colour centres in
  chemical vapour deposition nano-diamonds on iridium.
\newblock {\em New J. Phys.}, 13(025012), 2011.

\bibitem{Burek12}
M.~J. Burek, N.~P. de~Leon, B.~J. Shields, B.~J.~M. Hausmann, Y.~Chu, Q.~Quan,
  A.~S. Zibrov, H.~Park, M.~D. Lukin, and M.~Loncar.
\newblock Free-standing mechanical and photonic nanostructures in
  single-crystal diamond.
\newblock {\em Nano Lett.}, 12:6084--6089, 2012.

\bibitem{Hillery09}
M.~Hillery.
\newblock An introduction to the quantum theory of nonlinear optics.
\newblock {\em Acta Physica Slovaca}, 1:1--80, 2009.

\bibitem{Johnson2001:mpb}
Steven~G. Johnson and J.~D. Joannopoulos.
\newblock Block-iterative frequency-domain methods for {M}axwell's equations in
  a planewave basis.
\newblock {\em Opt. Express}, 8(3):173--190, 2001.

\end{thebibliography}

\section{Introduction}

Solid-state single photon emitters are an important step towards
scalable quantum information technology
\cite{Michler00,Pelton02,Duan04}. A single emitter resonantly coupled to a
high-finesse microcavity offers a natural functional unit to
realize important applications in quantum cryptography and
communication sciences \cite{Cirac97,Kuhn02,Childress06,Kimble08}. In
recent years, there has been strong interest in  
the negatively-charged nitrogen-vacancy ($\mathrm{NV}^-$) color center 
in diamond for coherent control of its optical and spin properties\cite{Kurtsiefer00,Beveratos02,Babinec10}.
However, although the NV center has long spin coherence times, of importance for
many quantum applications \cite{Jelezko04,Prawer08,Dutt07,Neumann08}, its fluorescence spectrum
is stretched over a huge incoherent phonon side-band, that constitues 96\% of the entire emission~\cite{McCutcheon08}. More recently, 
other color centers in diamond were explored as a possible alternative to NV~\cite{Prawer09,Prawer11}; for example, 
the ZPL of silicon-vacancy (SiV) center (ZPL $\sim 738~\mathrm{nm}$) is found to be much stronger than that of NV, constituting 70\% of the spectrum  
\cite{Jelezko13}. Recent studies have also established
the fundamental attributes of the SiV emitter including its electronic
structure and polarization states~\cite{Benson14,Becher12,Becher14,Jelezko13} as well as single photon indistinguishability from these sources~\cite{Sipahigil14},
validating SiV as a strong candidate for chip-based quantum optical applications.

On the fabrication side, many important steps have been taken to integrate diamond color centers
into chip-scale devices, including hybrid approaches \cite{Englund10,Wolters10,Sar11,Leon12}
in which diamond samples with color centers are juxtaposed to cavities fabricated in non-diamond
materials, as well as monolithic all-diamond approaches
\cite{Faraon11,Hausmann12,Faraon12,Moeller12,Burek12OM,Hausmann13NL,Burek14} in which the
cavity itself is hewn out of single crystal diamond. The
latter, though technically more challenging, has proven to be a
superior platform for diamond-based quantum photonics. On the other hand,
 the utility of third-order nonlinearity \ch in diamond
microphotonic devices has been explored in a recent work
\cite{Hausmann14} which demonstrates the generation of frequency combs
by parametric oscillators fabricated in single crystal diamond thin
films. Such capabilities open up an intriguing possibility -- to build
a monolithic all-diamond emitter/frequency-converter in a single
cavity design that efficiently collects and down-converts the color-center
photon into low-loss telecom frequency channels for long distance
communication. 

Chipscale frequency down-conversion of quantum signals has been
discussed by various authors for different material systems
\cite{McCutcheon09,Agha12,Huang13}. In particular, Ref.~\cite{Huang13}
has proposed a silicon nitride (SiN) micro-ring resonator that can
convert {\it few-photons} coherent light states from visible to
telecom frequencies. Here, we propose a monolithic diamond structure that can be
used to convert color-center {\it Fock-state single} photons from emission 
wavelengths to telecommunication bands ($\sim 1.5~\mathrm{\mu m}$).
We present a detailed theoretical description of the
conversion process, analyzing important practical concerns such as
nonlinear phase shifts and frequency mismatch, which were not
considered in previous papers. Additionally, we present an efficient
design technique for realizing perfect phase-matching in nonlinear
optical cavities. Our analysis predicts sustainable power requirements
for a chipscale nonlinear device with high conversion efficiencies
achievable at total pump powers below one Watt. 

\begin{figure}[t!]
\centering
\includegraphics[scale=0.5]{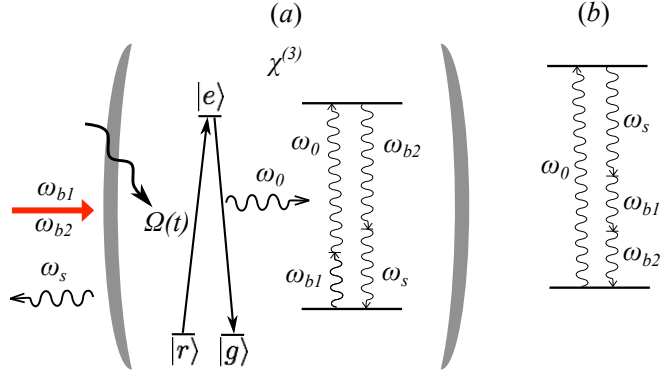}
\caption{(a) Schematic of an
  emitter-cavity system in which a single emitter is embedded in a nonlinear
  \ch cavity supporting four resonant modes at frequencies
  $\{\omega_{0c},~\omega_{sc},~\omega_{b1c},~\omega_{b2c}\}$.
  The eigenstructure of the emitter is represented by a simplified
  three-level system with states $\ket{r},~\ket{e},~\ket{g}$. A laser
  trigger with frequency $\omega_{re}$ and intensity $\propto
  |\Omega|^2$ addresses the emitter states $\ket{r}$ and $\ket{e}$.
  The photon released from the emitter is collected by the cavity and 
  down-converted to telecom through the
  four-wave mixing Bragg scattering process (FWM-BS).
  The latter process obeys the frequency-matching
  relation $\omega_s + \omega_{b2} = \omega_0 + \omega_{b1}$, where
  $\omega_0$ and $\omega_s$ are the frequencies of the emitter and telecom
  signals, and $\omega_{b1}$ and $\omega_{b2}$ are the frequencies of
  the NIR and telecom pump lasers respectively. (b) Alternatively, the single photon ($\omega_0 $) can be down-converted by 
  the difference frequency generation (DFG) process
  in which it is broken up
  into one signal ($\omega_s$) and two pump
  photons ($\omega_{b1},~ \omega_{b2}$), satisfying the 
  frequency relation: 
  $\omega_0 - \omega_{b1} - \omega_{b2} = \omega_s$. 
  Here, we choose the two pump photons to be degenerate, 
  $\omega_{b1}=\omega_{b2}=\omega_b$.\label{fig:fig1}}
\end{figure}

Although our theoretical analysis is generally valid for \emph{any} color center,
we primarily tailor our cavity designs to the SiV emitter. 
Specifically, we consider a single SiV emitter implanted in 
a high-Q microcavity. In the case of triggered single photon emission, 
the electronic structure of the emitter
can be well-approximated by a simple three-level diagram \cite{McCutcheon09} 
with the electronic states denoted by $\ket{r},~\ket{e}$ and $\ket{g}$ (\figref{fig1}a).
To start with, the emitter is prepared in the state $\ket{r}$. 
An external trigger excites the system to the state $\ket{e}$ which
subsequently decays to the ground $\ket{g}$, emitting a
photon of frequency $\omega_0$ into the surrounding cavity.
The photon release is followed by frequency down-conversion (within the same cavity) via
a nonlinear wave-mixing process which utilizes the inherent third order susceptibility \ch of diamond.

A common scheme for \ch-based frequency down-conversion is four-wave mixing Bragg-scattering
(FWM-BS) \cite{Agha12,Huang13}, in which the incoming photon at emitter frequency $\omega_0$
is scattered by coherent pump lasers at frequencies $\omega_{b1}$ and $\omega_{b2}$ to
yield an outgoing photon at frequency $\omega_s$, satisfying the frequency matching condition
$\omega_0 + \omega_{b1}= \omega_s + \omega_{b2}$ (\figref{fig1}a). 
Alternatively, the down-conversion can be realized by a difference frequency generation (DFG) scheme
in which the SiV emitter photon is directly broken up
 into one signal and two pump photons, with frequencies satisfying the relation $\omega_0 - \omega_{b2} - \omega_{b1}= \omega_s $ (\figref{fig1}b). 
In this paper, we present carefully designed diamond microcavities that can enhance frequency down-conversion
via FWM-BS or DFG. In particular, we design our cavities to host four resonant modes $(\omega_{0c},~\omega_{sc},~\omega_{b1c},~\omega_{b2c})$
that respect the frequency matching conditions,
$\omega_{0c} + \omega_{b1c} \approx \omega_{sc} + \omega_{b2c} $ for FWM-BS and
$\omega_{0c} - \omega_{b1c} - \omega_{b2c} \approx \omega_{sc} $ for DFG. 

In what follows, we mainly focus our discussion on FWM-BS; 
however, the analysis is equally applicable to DFG, for which we briefly discuss design specifications, efficiency and power predictions.
In our analysis, we take into account the effects of frequency mismatch as well as those of self and cross-phase modulation 
(ignored in the earlier work \cite{Huang13}) and find that even in
their presence, maximum efficiency can be achieved by appropriate power inputs. 
We found that
the maximum efficiencies can vary from 30 \% to 90\%, depending on the 
internal quantum efficiency of the color center (i.e., the ratio of radiative to non-radiative decay) 
and the quality factor of the cavity mode at the emitter frequency. Our cavity designs 
ensure that these efficiencies can be achieved by 
input pump powers lower than one Watt.

\section{Theoretical considerations}

The effective Hamiltonian of the emitter-cavity system including the dissipative terms has the following form (in the Schr\"{o}dinger
picture):
\begin{align}
  \mathcal{H} &= \hbar \omega_{0c} ~ \hat{a}_0^{\dagger}\hat{a}_0 
	       + \hbar \omega_{sc} ~ \hat{a}_s^{\dagger}\hat{a}_s 
	       + \hbar \omega_0 ~    \hat{\sigma}_{ee}
    	       + \hbar \omega_r ~ \hat{\sigma}_{rr} \notag \\
    & - \sum_{\mu, b} \hbar \omega_{\mu c} \alpha_{\mu b} |a_b|^2~ \hat{a}_{\mu}^{\dagger}\hat{a}_{\mu} \notag \\
    & + \sum_l \hbar \omega_{lc} \left(1 - \alpha_{lb} |a_b|^2 \right) ~ \hat{a}_l^{\dagger}\hat{a}_l \notag \\
    & + \hbar \Omega(t) e^{-i \omega_{re} t} \hat{\sigma}_{er}
    + \hbar \Omega^*(t) e^{i \omega_{re} t} \hat{\sigma}_{re} \notag \\
    & + \hbar g_\text{ZPL} \hat{\sigma}_{eg} \hat{a}_0
    + \hbar g_\text{ZPL} \hat{\sigma}_{ge} \hat{a}_0^{\dagger} \label{eq:Ham} \\
    & - \hbar \beta ~ \hat{a}_s^{\dagger}\hat{a}_0 e^{i (\omega_{b2} -
      \omega_{b1}) t}
    - \hbar \beta^* ~ \hat{a}_0^{\dagger}\hat{a}_s e^{-i (\omega_{b2} - \omega_{b1}) t} \notag \\
    & + \hbar \sum_l \left( g_l \hat{\sigma}_{eg} \hat{a}_l
      + 		    g_l \hat{\sigma}_{ge} \hat{a}_l^{\dagger} \right) \notag \\
    & - i \hbar {\kappa_0 \over 2} ~ \hat{a}_0^{\dagger}\hat{a}_0 - i
    \hbar { \kappa_s \over 2} ~ \hat{a}_s^{\dagger}\hat{a}_s - i \hbar
    \sum_l \left( {\kappa_l \over 2} ~ \hat{a}_l^{\dagger}\hat{a}_l
    \right) - i \hbar { \gamma^\text{NC} \over 2} ~
    \hat{\sigma}_{ee}. \notag
\end{align}

Here, $\hat{a}$ and $\hat{a}^\dagger$ are annihilation and creation
operators for the quantized cavity field at the emitter $(0)$ and telecom $(s)$
wavelengths, as well as at certain other wavelengths $(l)$ that happen to fall within the
phonon sidebands of the emission spectrum. The modal amplitudes of
the classical laser pumps are denoted by $a_b$ ($b \in \{b1,b2\}$) and
are normalized so that $|a_b|^2$ is the energy in the $b$
mode. $\hat{\sigma}_{ij}=\ket{i}\bra{j}$ is the atomic operator
connecting the $j$th and $i$th emitter states. $\Omega(t)$ is the 
Rabi frequency coupling the trigger pump and $r \leftrightarrow e$ 
transition whereas $g_\text{ZPL}$ addresses the coupling
between the ZPL decay and the 0th cavity mode.
Similarly, the coefficients $g_l$ describe the phonon sideband
transitions that coincide with the cavity resonances at wavelengths
$l$. Our model also includes self and cross-phase modulation via coefficients $\alpha$
introduced by the presence of the classical pump fields, whereas the strength 
of frequency conversion is described by $\beta$. These nonlinear coupling coefficients
can be calculated from perturbation theory (see
Appendix~\ref{hamiltonian}). The leakage of each cavity
mode is denoted by $\kappa$, which is related to the quality factor of the mode by
$\kappa = {\omega_c \over Q}$. $\gamma^\text{NC}$ denotes the
spontaneous decay rate of the emitter into continuum loss channels, including
both radiative and non-radiative decays \cite{McCutcheon09}. Note that in the above Hamiltonian, we
have omitted nonlinear phase-modulation effects $ \sim \left( \hat{a}^\dagger \right)^2 \hat{a}^2 $ \cite{Drummond90} owing to the
quantum signals themselves, as these effects are vanishingly small at the single-photon level.

In addition to the Hamiltonian describing the quantum-mechanical
degrees of freedom, we must also consider the coupled-mode equations
for the classical degrees of freedom, namely the pump 
amplitudes $a_{b1}$ and $a_{b2}$. These equations are given
by~\cite{Ramirez11}:
\begin{align}
  {da_{b1} \over dt} &= i \omega_{b1c} \left( 1 - \alpha_{bb1}
    |a_{b1}|^2 - \alpha_{b1b2} |a_{b2}|^2 \right) a_{b1}
  - {a_{b1} \over \tau_{b1}} + \sqrt{{2 \over \tau_{sb1}} P_{b1}},  \notag \\
  {da_{b2} \over dt} &= i \omega_{b2c} \left( 1 - \alpha_{bb2}
    |a_{b2}|^2 - \alpha_{b1b2} |a_{b1}|^2 \right) a_{b2} - {a_{b2}
    \over \tau_{b2}} + \sqrt{{2 \over \tau_{sb2}} P_{b2}},  \notag
\end{align}
where $P_{b1}$ and $P_{b2}$ denote the powers of the incoming lasers
coupling to the cavity modes $\omega_{b1c}$ and $\omega_{b2c}$, with
corresponding coupling lifetimes $\tau_{sb1}={2 Q_{sb1} \over
  \omega_{b1c}}$ and $\tau_{sb2}={2 Q_{sb2} \over
  \omega_{b2c}}$. Here, $\tau_{b1}={2 Q_{sb1} \over \omega_{b1c}}$ and
$\tau_{b2}={2 Q_{b2} \over \omega_{b2c}}$ are the overall lifetimes of
the two cavity modes, which include all other sources of cavity losses
(e.g. radiation and/or material absorption). Since one has the freedom
to choose the frequencies $(\omega_{b1},~\omega_{b2})$ of the incoming
pump lasers, it is always possible to operate under the resonant
condition for the classical pumps,
\begin{align}
  \omega_{b1} &= \omega_{b1c} \left( 1 - \alpha_{bb1} |a_{b1}|^2 - \alpha_{b1b2} |a_{b2}|^2 \right) \label{eq:classres1} \\
  \omega_{b2} &= \omega_{b2c} \left( 1 - \alpha_{bb2} |a_{b2}|^2 - \alpha_{b1b2} |a_{b1}|^2 \right), \label{eq:classres2}
\end{align}
which allows us to ignore effects associated with self- and
cross-phase modulation~\cite{Rodriguez07:OE,Ramirez11}. It follows that the steady-state energy
in the classical cavity mode are given by $|a_{b1}|^2 = {2\tau_{b1}^2
  \over \tau_{sb1}} P_{b1},~|a_{b2}|^2 = {2\tau_{b2}^2 \over
  \tau_{sb2}} P_{b2}$. 
Linear stability analysis \cite{Rodriguez07:OE,Ramirez11} also confirms that
these solutions are stable. 

The nonlinear coupling coefficients can be calculated from the
perturbation theory \cite{Rodriguez07:OE,Johnson01:cavities} (see also
Appendix~\ref{hamiltonian}):
\begin{align}
  \beta =& {3 \over 4} ~\sqrt{\omega_{0c} \omega_{sc} ({2 \tau_{b1}^2 \over \tau_{sb1}} {2 \tau_{b2}^2 \over \tau_{sb2}}) P_{b1} P_{b2} } \notag \\
  & \frac{\int dV ~\epsilon_0 ~ 
    \{  \chi_{xxyy}~ (\vE_s^* \cdot \vE_0 )     ( \vE_{b1} \cdot \vE_{b2}^* )
    +  \chi_{xyxy}~ (\vE_s^* \cdot \vE_{b1}) ( \vE_0 \cdot \vE_{b2}^* ) 
    +  \chi_{xyyx}~ (\vE_s^* \cdot \vE_{b2}^*) ( \vE_0 \cdot \vE_{b1} ) \} }
  {\sqrt{\int dV ~ \epsilon_0 |\vE_0|^2}~\sqrt{\int dV \epsilon_s |\vE_s|^2}
    ~\sqrt{\int dV ~ \epsilon_{b1} |\vE_{b1}|^2}~\sqrt{\int dV ~ \epsilon_{b1} |\vE_{b2}|^2}  }, \label{eq:beta}\\
  \alpha_{\mu b} =& {3 \over 4} ~
  \frac{\int dV ~\epsilon_0 
    \{   \chi_{xxyy}~ |\vE_{\mu}|^2 |\vE_b|^2  
    + \chi_{xyxy}~ | \vE_{\mu} \cdot \vE_b^* |^2
    + \chi_{xyyx}~ |\vE_{\mu} \cdot \vE_b |^2 \}}
  { (\int dV ~ \epsilon_0 |\vE_{\mu}|^2) (\int dV ~ \epsilon_b |\vE_b|^2)} , \label{eq:aub}\\
  \alpha_{bb} =& { 3 \over 8} \frac{\int dV ~ \epsilon_0~
    \{ (\chi_{xyxy} + \chi_{xyyx})~ |\vE_b \cdot \vE_b^*|^2 
    + \chi_{xxyy}~ |\vE_b \cdot \vE_b|^2 \} }
  {\left( \int dV ~ \epsilon_b |\vE_b|^2 \right)^2}, \label{eq:abb}\\
  \alpha_{b1b2} =& {3 \over 4} \frac{\int dV ~ \epsilon_0~ 
    \{ \chi_{xxyy}~|\vE_{b1} \cdot \vE^*_{b2}|^2 
    + \chi_{xyxy}~|\vE_{b1} \cdot \vE_{b2}|^2 
    + \chi_{xyyx}~|\vE_{b1}|^2 |\vE_{b2}|^2 \} }
  {\left( \int dV ~ \epsilon_{b1} |\vE_{b1}|^2 \right) 
    \left( \int dV ~ \epsilon_{b2} |\vE_{b2}|^2 \right)}. \label{eq:ab1b2}
\end{align}  

The initial state of the emitter-cavity system is simply a tensor product of 
the emitter state $\ket{r}$ and the vacuum photonic states $\ket{0_0},~\ket{0_s}$ and $\ket{0_l}$:
\begin{align}
\ket{\Psi(0)} =  \ket{r,0_0,0_s,0_l}.
\end{align}
We now proceed to solve the time evolution of $\Psi$ under the Hamiltonian \eqref{Ham}.
It is easy to see that the initial state $\ket{r,0_0,0_s,0_l}$ can only couple to states 
with one quantum of excitation, $\ket{e,0_0,0_s,0_l}$ or $\ket{g,n_0,n_s,n_l},~n_0+n_s+n_l=1$.
Therefore, it is sufficient to expand the total wavefunction as a superposition of 
these \emph{single-excitation} states~\cite{McCutcheon09}:
\begin{align}
\ket{\Psi(t)} &=   c_r(t) e^{- i \omega_r t} \ket{r,0_0,0_s,0_l} \notag \\
		&+ c_e(t) e^{- i \omega_0 t} \ket{e,0_0,0_s,0_l} \notag \\
		&+ c_0(t) e^{- i \omega_0 t} \ket{g,1_0,0_s,0_l} \label{eq:wvf} \\
		&+ c_s(t) e^{- i \omega_s t} \ket{g,0_0,1_s,0_l} \notag \\
		&+ \sum_l c_l(t) e^{- i \omega_l t} \ket{g,0_0,0_s,1_l}. \notag
\end{align} 
Substituting \eqref{wvf} in the time-dependent Schr\"{o}dinger
equation, $ i \hbar {\partial \ket{\Psi} \over \partial t} = \mathcal{H} \ket{\Psi}$, leads to the following coupled equations of motion for the
coefficients:
\begin{align}
  \dot{c}_r &= - i \Omega^*(t) c_e \label{eq:rate1}\\
  \dot{c}_e &= - i \Omega(t) c_r - i g_\text{ZPL} c_0 - i \sum_l g_l c_l - {\gamma^\text{NC} \over 2} c_e \\
  \dot{c}_0 &= - i \delta_{0c} c_0 - i g_\text{ZPL} c_e - i \beta^* c_s - {\kappa_0 \over 2} c_0 \\
  \dot{c}_s &= - i \delta_{sc} c_s - i \beta c_0 - {\kappa_s \over 2} c_s \\
  \dot{c}_l &= - i g_l c_e - {\kappa_l \over 2} c_l \label{eq:rate2}\\
  \delta_{0c} &= \omega_{0c} \left(1 - \alpha_{0b1} |a_{b1}|^2 - \alpha_{0b2} |a_{b2}|^2 \right) - \omega_0 \label{eq:d0}\\
  \delta_{sc} &= \omega_{sc} \left(1 - \alpha_{sb1} |a_{b1}|^2 - \alpha_{sb2} |a_{b2}|^2 \right) - \left( \omega_0 + \omega_{b1} - \omega_{b2} \right) \label{eq:ds}.
\end{align} 
The nonlinear detunings $\delta_{0c}$ and $\delta_{sc}$, as defined by \eqreftwo{d0}{ds},
include the nonlinear phase shifts introduced by the pumps and 
are to be distinguished from the
\emph{bare} cavity detunings $\omega_{0c} - \omega_0$ and 
$\omega_{sc} + \omega_{b2c} - \omega_{0c} - \omega_{b1c}$  \emph{without} the amplitude-dependent phase shifts.

\begin{figure}[t!]
\centering
\includegraphics[scale=0.6]{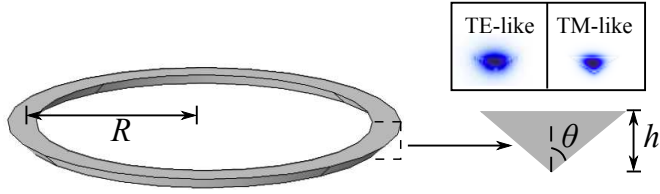}
\caption{Schematic of angle-etched diamond ring resonator (refractive
  index $n\approx 2.41$). The resonator has a radius $R$ and a height $h$. The
  triangular cross-section has an etch-angle $\theta$. 
  $E_r$ components of the fundamental TE0-like modes
  at frequencies $(\omega_{sc},~\omega_{b1c})$
  and $E_z$ components of the fundamental TM0-like modes at frequencies
  $(\omega_{0c},~\omega_{b2c})$ are also
  depicted in the picture. \label{fig:fig4}}
\end{figure}

In the regime where the trigger intensity $\sim \Omega$ and trigger rate $\sim {d \Omega \over dt}$ are
sufficiently small compared to the decay rates of the system, the
probability coefficients $c_{i},~i\in \{e,0,s,l\}$, follow the
dynamics of $c_r$. Thus we can adiabatically eliminate the $e,0,s,l$
degrees of freedom (see Appendix~\ref{adiabatic-elimination}), and
arrive at the following expression for the conversion efficiency
(defined as the percentage of triggers for which the excitations are ultimately down-converted to telecom \cite{McCutcheon09}):
\begin{align}
  F &= \frac{ \int dt ~\kappa_{ss} |c_s|^2 } { \int dt ~ \left( \kappa_s |c_s|^2 + \kappa_0 |c_0|^2 + \sum_l \kappa_l |c_l|^2 + \gamma^\text{NC} |c_e|^2 \right) } \\
  &= \frac{4 C |\beta|^2 \kappa_0 \kappa_{ss} } {16 |\beta|^4 + 4
    |\beta|^2 \left( (2 + C) \kappa_0 \kappa_s - 8 \delta_{0c}
      \delta_{sc} \right) + \left( 4 \delta_{0c}^2 + (1 + C)
      \kappa_0^2 \right) (4 \delta_{sc}^2 +
    \kappa_s^2)} \label{eq:eff}.
\end{align}
Here, $C$ is the well-known cooperativity factor:
\begin{align}
  C = { 4 g^2_\text{ZPL} \over \kappa_0 \left( \sum_l {4g^2_l \over
        \kappa_l} + {\gamma^\text{NC}} \right) } \label{eq:cop},
\end{align}
which describes the {\it relative} strength of photon capture into the cavity 0th mode 
over the competing loss channels. \eqref{eff} can be maximized by
setting ${\partial F
  \over \partial \delta_{0c}} = 0,~{\partial F \over \partial
  \delta_{sc}} = 0$, which yields:
\begin{align}
  F^\text{max} = \left(1+ {2 \over C} - 2 \sqrt{{1 \over C} + {1 \over
      C^2}}\right)
~{Q_s \over Q_{ss}} \label{eq:maxeff},
\end{align}
under the constraints,
\begin{align}
  \delta_{0c}^\text{crit} = \delta_{sc}^\text{crit} = \pm {1 \over 2} \sqrt{ {\kappa_0 \over
      \kappa_s} \left( 4 \sqrt{1 + C} |\beta|^2 - \left(1 + C\right)
      \kappa_0 \kappa_s \right) }\label{eq:dcrit}.
\end{align}
A special situation arises when 
\begin{align}
\delta_{0c}^\text{crit}=\delta_{sc}^\text{crit}=0, \label{eq:dcrit0}
\end{align}
in which case $\beta$ is constrained by,
\begin{align}
  |\beta|^\text{crit} = { \left( 1 + C \right)^{1/4} \over 2 }~ \sqrt{
    {\omega_{0c} \omega_{sc} \over Q_0 Q_s} }. \label{eq:bcrit0}
\end{align}
This scenario was already discussed in Ref.~\cite{McCutcheon09}. More
generally, we can algebraically solve \eqref{dcrit} together with
\eqreftwo{classres1}{classres2}, to yield a set of critical powers and frequencies
$\{\omega_{b1}^\text{crit},~P_{b1}^\text{crit},~\omega_{b2}^\text{crit},~P_{b2}^\text{crit}\}$.
However, it must be noted that the above algebraic equations sometimes lead to negative or imaginary answers.
Such cases indicate that the maximal efficiency is no longer given by \eqref{maxeff};
instead the critical efficiencies and powers are numerically found by directly setting 
\begin{align}
{\partial F \over \partial P_{b1}} = 0,\quad {\partial F \over \partial P_{b2}} = 0. \label{eq:optP}
\end{align} 
In the following sections, we will perform
these calculations in the context of carefully designed realistic cavity systems.

\section{Diamond Micro-ring Resonator}
\label{secDesign}

\begin{figure}[t!]
\centering
\includegraphics[scale=0.5]{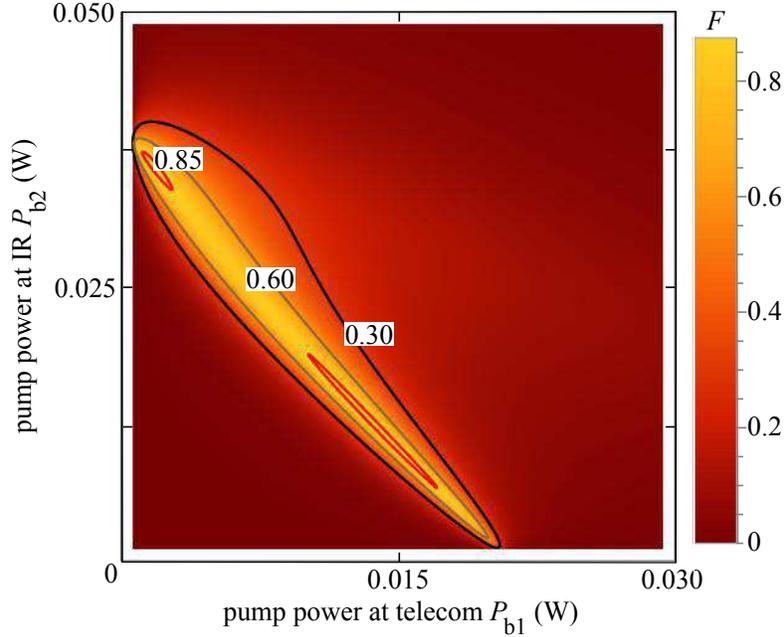}
\caption{Density plot of conversion efficiency $F$
  (defined in \eqref{eff}) over the pump powers $P_{b1}$ and $P_{b2}$
  for the cavity system discussed in Section~\ref{secDesign}.
  Efficiency contours are overlaid on the plot for easy visualization.
  They help identify the regime of pump powers necessary for high-efficiency conversion.
  \label{fig:fig2}}
\end{figure}

\begin{figure}[h!]
\centering
\includegraphics[scale=0.5]{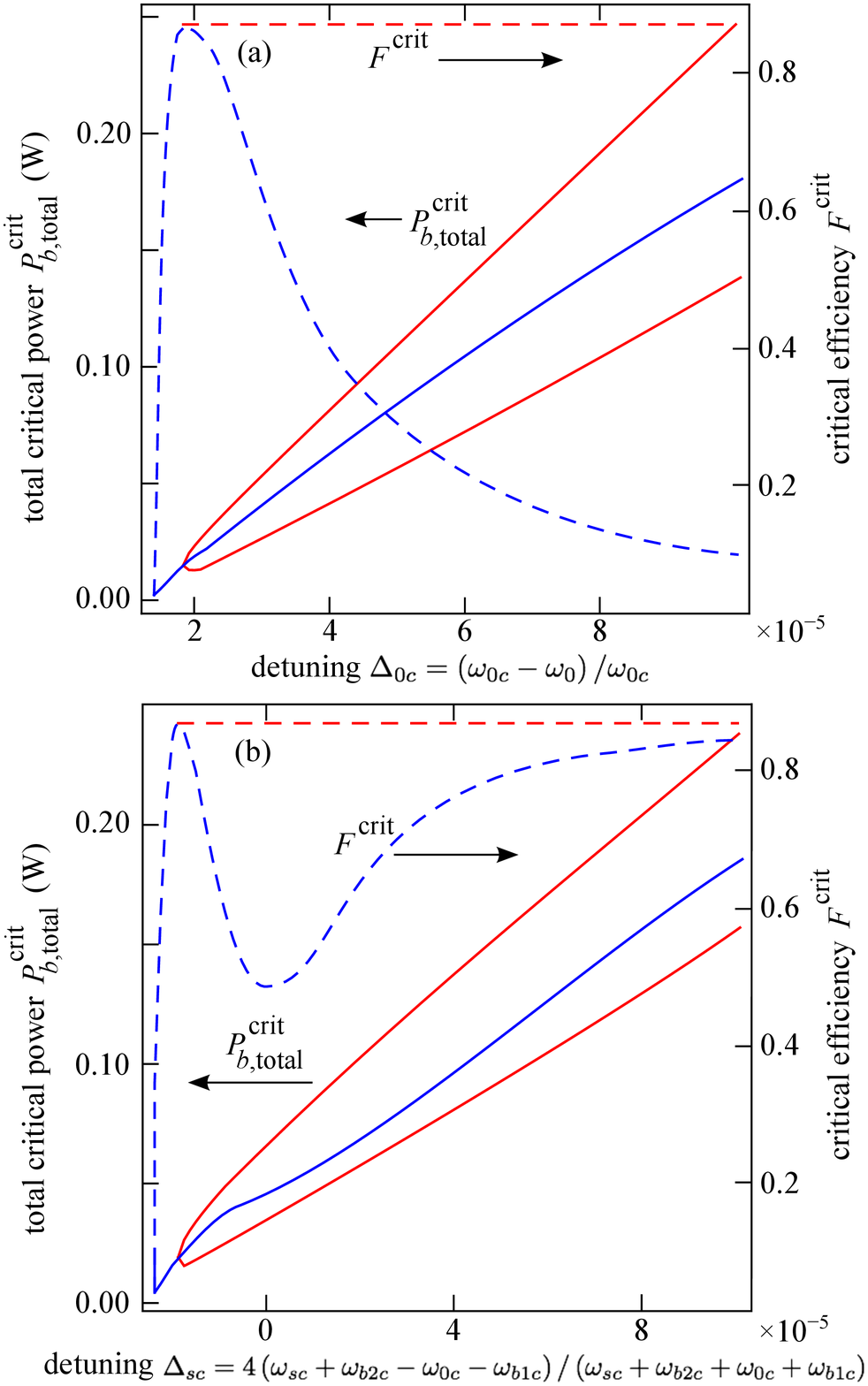}
\caption{(a) Critical pump power (solid lines, left-axis) 
and critical efficiency (dashed lines, right-axis)
vs. bare-cavity fractional detuning 
$\Delta_{0c} = 1 -\omega_0/\omega_{0c}$.
Pump power plotted here is the total pump power,
$P_{b,\text{total}}^\text{crit} =
P_{b1}^\text{crit} + P_{b2}^\text{crit}$.  
Two regimes can be identified: a regime with multiple critical
points (light red and dark blue colors) and a regime with 
only one critial point (dark blue only). The former corresponds
to the regime where \eqref{maxeff} is valid and the detuning
can be compensated by critical pump powers.
The latter corresponds to the regime where \eqref{maxeff} is no longer valid and
the efficiency falls off rapidly with detuning. (See also text.)
(b) same as above except for the detuning 
$\Delta_{sc}= { \omega_{sc} + \omega_{b2c} -
  \omega_{0c} -\omega_{sc} \over \left(\omega_{sc} + \omega_{b2c} +
    \omega_{0c} + \omega_{b1c} \right)/4 }$.
\label{fig:fig3}}
\end{figure}

In this section, we consider concrete and realistic cavity designs
for single photon frequency conversion based on a triangular ring resonator.
We choose the triangular cross-section to take advantage of our recently developed angle-etched technique on monolithic bulk diamond~\cite{Burek14}. 
The main advantage of angle-etched technique is its consistent yield and 
scalability compared to diamond-on-insulator thin-film techniques.

The most challenging task in our cavity design is to satisfy the 
frequency-matching relation $\omega_{0c} + \omega_{b1c} \approx
\omega_{sc} + \omega_{b2c}$. One cannot straightforwardly apply 
conventional dispersion engineering techniques developed for FWM with closely-spaced frequencies~\cite{Hansryd02} 
since, in our case, the conversion spans two very different frequency bands, NIR and telecom.
One can do an extensive numerical search in the design space \cite{Huang13} but
such an approach is time-consuming and computationally very expensive.
In contrast, in the Appendix~\ref{design}, we describe
a numerical/visualization technique that greatly simplifies the
process of designing ring resonators that satisfy the frequency-matching condition. One
such cavity is shown in \figref{fig4}. The ring has a radius $R = 14.4~\mathrm{\mu m}$, 
a height $h = 414~\mathrm{nm}$ and a slant angle $\theta = 59^\circ$.
Out of its many resonances, we will focus our
attention on two fundamental transverse-electric-like ($\mathrm{TE}_0$-like) modes in the telecom band, 
$\left(m_s=95,\lambda_{sc} = 1.613~\mathrm{\mu m}\right)$ and $\left(m_{b1}=79,\lambda_{b1c}=1.806~\mathrm{\mu m} \right)$,
 and two fundamental transverse-magnetic-like ($\mathrm{TM}_0$-like) modes in the NIR band, $\left(m_0=262,\lambda_{0c}=738~\mathrm{nm}\right)$ and
$\left(m_{b2}=246,\lambda_{b2c}=776~\mathrm{nm} \right)$, see
\figref{fig4}. Finite-difference time-domain (FDTD) simulations reveal that the four modes have
ultra-high radiative quality factors in excess of $10^8$.
However, in actual experiments, the overall $Q$'s will be limited by fabrication imperfections or coupling losses.
Therefore, we will examine the maximum efficiencies for two different representative cases, $Q\sim10^5$ and $Q\sim10^6$.
For the sake of simplicity, we will assume that the overall $Q$'s are
coupling-limited, so that $Q_i \approx Q_{si},~i\in \{0,s,b1,b2\}$. If
this is not the case, the maximal efficiency will suffer by a
factor of $Q_s/Q_{ss}$ (see \eqref{maxeff}). The third-order nonlinear 
susceptibility of diamond (Kerr coefficient) is $\ch = 2.5 \times 10^{-21} {\mathrm{m}^2\over \mathrm{V}^2}$ \cite{Boyd92}.
The nonlinear phase-shifts $\alpha$ and coupling coefficient $\beta$ are found to be:
\begin{align}
  &|\beta| = 0.120~\mathrm{(\mu J)^{-1}} ~ \sqrt{\omega_{0c}
    \omega_{sc} \left({2 \tau_{b1}^2 \over \tau_{sb1}} {2 \tau_{b2}^2
        \over \tau_{sb2}}\right) P_{b1} P_{b2} }
  \\
  &\alpha_{0b1} = 0.114~\mathrm{(\mu J)^{-1}},
  \quad 
  \alpha_{0b2} = 0.425~\mathrm{(\mu J)^{-1}},
  \\ 
  &\alpha_{sb1} = 0.215~\mathrm{(\mu J)^{-1}},
  \quad 
  \alpha_{sb2} = 0.126~\mathrm{(\mu J)^{-1}},
  \\ 
  &\alpha_{bb1} = 0.099~\mathrm{(\mu J)^{-1}},
  \quad 
  \alpha_{bb2} = 0.208~\mathrm{(\mu J)^{-1}}, \quad
  \\ 
  &\alpha_{b1b2} = 0.114~\mathrm{(\mu
    J)^{-1}} .
\end{align}

The maximum acheivable efficiency (\eqref{maxeff}) is determined by the
cooperativity factor $C$ (\eqref{cop}), which can be recast as:
\begin{align}                                                                                 
  C = { 4 g^2_\text{ZPL} \over \kappa_0 \gamma } ~ \left( \sum_l
    {4g^2_l \over \kappa_l \gamma} + {\gamma^\text{NC} \over \gamma}
  \right)^{-1},
\label{eq:C}
\end{align}
where $\gamma$ is the total spontaneous decay rate of the color center in
the isotropic diamond medium in the absence of the cavity. The
quantity $\sum_l {4g^2_l \over \kappa_l \gamma}$ represents the relative strength of radiative decay
into phonon sidebands in the presence of the ring resonator as opposed to
the total decay rate in the absence of the resonator.  Generally, one 
can compute this quantity by simulating the decay rates
of all of the ring resonator modes falling within the phonon
sidebands. However, in the case of SiV with a relatively weak phonon 
sideband, we can expect much of the sideband to be  
suppressed in between the ring resonator modes so that $\sum_j {4g^2_j \over \kappa_j \gamma} \ll 1$. 
Therefore, it follows that
\begin{align}
&\sum_l {4g^2_l \over \kappa_l \gamma} + {\gamma^\text{NC} \over \gamma} \leq 1 \quad\text{or}\\
 & C \geq { 4 g^2_\text{ZPL} \over \kappa_0 \gamma }
\end{align}
since ${\gamma^\text{NC} \over \gamma} < 1$ in general. In our calculations, we will 
simply take the lower bound $C = { 4 g^2_\text{ZPL} \over \kappa_0 \gamma }$.
Assuming that the SiV is located at the electric-field maximum and that the
dipole element is aligned with the field polarization, the coupling
strength is given by $g_\text{ZPL} = {\mu_\text{ZPL}
  |\mathbf{E}_\text{max}| \over \hbar}$. Spontaneous emission rate $\gamma_\text{ZPL}$ is related to
the dipole element via $\gamma_\text{ZPL} = {n \omega_0^3
  \mu_\text{ZPL}^2 \over 3 \pi \epsilon_0 \hbar c^3}$. The mode volume
of the 0th cavity mode is $V_{0c} = { \int~dV \epsilon |\mathbf{E}|^2 \over
  \epsilon(\mathbf{r}_\text{max}) |\mathbf{E}_\text{max}|^2 } =
252~{\left(\lambda \over n\right)}^3$. The branching ratio of the SiV
ZPL line is ${\gamma_\text{ZPL} \over \gamma^\text{rad}} \approx 70\%$
\cite{Jelezko13}. A variable factor in our calculations is the internal quantum efficiency 
of the SiV emission (or the ratio $\gamma^\text{rad} /\gamma$); the current literature 
suggests it could vary wildly from sample to sample, from as low as $9\%$ \cite{Neu11} to as high as near $100\%$ at low temperatures~\cite{Becher14}, 
probably depending on material quality and method of fabrication. In Table.~\ref{tab1}, we have computed
maximal efficiencies and critical pump powers (\eqref{maxeff} and \eqreftwo{dcrit0}{bcrit0}) for different internal quantum efficiencies and cavity lifetimes. A maximal conversion efficiency of 87\% in the best possible scenario where $\gamma^\text{rad} /\gamma=1$ and $Q=10^6$ is limited by a moderate cooperativity $C \approx 211$ (see Eq.~\eqref{maxeff}). However, we note that this limit only reflects the current experimental constraints on diamond fabrication ($Q\sim 10^6$); simulations have indicated that radiative quality factors can be in excess of $10^7$, in which case the maximum efficiency quickly approaches 100\%.

\begin{table}
\centering
\begin{tabular}{|c|c|c|c|c|c|c|c|c|c|c|c|c|}
  \hline
  $Q$  	 & $\gamma^\text{rad} /\gamma$    &$C$ 	& $P_{b1}^\text{crit} ~(\mathrm{W})$ 	& $P_{b2}^\text{crit}~(\mathrm{W})$ 	     	& $F^\text{max}$ \\ \hline
  $10^5$ & 0.09	     	    		  &2	& 0.110    				& 0.424    	     				& 0.26  	 \\ \hline
  $10^5$ & 1	     	    		  &21	& 0.255    				& 0.507		     				& 0.65  	 \\ \hline
  $10^6$ & 0.09	     	    		  &19	& $3\times 10^{-4}$    			& 0.036						& 0.63  	 \\ \hline
  $10^6$ & 1	     	    		  &211	& 0.001		   			& 0.037		 				& 0.87  	 \\ \hline
  \hline
\end{tabular}
\caption{Critical powers and efficiencies. \label{tab1}}
\end{table}

Next, we examine the scenario with $Q=10^6$ and ${\gamma^\text{rad} \over \gamma}=1$
as the best possible performance for the given design.
\Figref{fig2} shows a density plot of the conversion efficiency
(\eqref{eff}) as a function of the two pump powers.
As observed in the figure, there exist narrow strips of high-efficiency 
regime $(F>0.85)$ around the pump power co-ordinates (1~$\mathrm{mW}$, 35~$\mathrm{mW}$) and (15~$\mathrm{mW}$, 12~$\mathrm{mW}$). 
A closer examination reveals two additional critical
points (in addition to the one given in Table.~\ref{tab1}) with efficiencies 
$F^\text{crit,1}=0.87,~F^\text{crit,2}=0.82$ 
and pump powers $P_{b1}^\text{crit,1} \approx 14~\mathrm{mW},~P_{b1}^\text{crit,2} \approx 11~\mathrm{mW}$ 
and $P_{b2}^\text{crit,1} \approx 5~\mathrm{mW},~P_{b2}^\text{crit,2} \approx 28~\mathrm{mW}$ respectively.
In contrast to the tabulated critical point, these extra points have a small but non-vanishing phase mismatch $\delta \neq 0$
and are obtained from a complete solution of the algebraic equation (\eqref{dcrit}) or from the direct optimization over pump powers (\eqref{optP}). 
The existence of multiple
critical points is a result of a complex interplay between two competing
decay channels, (i) the decay of the excited state with the
concomitant photon release into mode $0$ and (ii) the conversion of
the released photon to the telecom mode $s$. If the leakage from mode
$0$ to $s$ is too strong, no cavity enhanced emission occurs,
degrading the efficiency, whereas if this leakage is too weak,
nonlinear conversion efficiency vanishes. These decay rates are
controlled by the nonlinear detunings, $\delta_{0c},~\delta_{sc}$, and
nonlinear coupling $\beta$ parameters, which are in turn controlled by
the input powers. Specifically, $\delta_{0c}$ tends to inhibit the
$\mathrm{emitter \rightarrow 0}$ decay channel, $\delta_{sc}$ tends to
inhibit the $0 \rightarrow s$ decay channel, and $\beta$ tends to
enhance the $0 \rightarrow s$ decay channel.  The complex interplay
between these parameters introduces multiple critical points over the
$P_{b1}$--$P_{b2}$ plane.

As we have shown above, the nonlinear detunings $\delta$'s are directly controlled 
by incident pump powers, which can, therefore, be appropriately chosen to 
ensure maximum conversion efficiency. The magnitudes of the critical pump
powers in turn depend on how much frequency detuning the system begins with. Theoretically,
the cavity parameters can be numerically optimized so that these detunings are made as small as possible. However,
random errors in the actual fabrication process can introduce significant deviations from any designated cavity
frequencies. Therefore, it is advisable to consider the effects of frequency detunings due to 
possible variations in ring radius $R$ and height $h$. For this purpose,
we define the \emph{bare}-cavity fractional detunings: $\Delta_{0c} = 1 -
\omega_0/\omega_{0c},~\Delta_{sc}= { \omega_{sc} + \omega_{b2c} -
  \omega_{0c} -\omega_{sc} \over \left(\omega_{sc} + \omega_{b2c} +
    \omega_{0c} + \omega_{b1c} \right)/4 }$.
\Figref{fig3} shows the {\it total} critical power $P_{b,\text{total}}^\text{crit} =
P_{b1}^\text{crit} + P_{b2}^\text{crit}$ and corresponding critical
efficiencies $F^\text{crit}$ as a function of $\Delta_{0c}$ (\figref{fig3}(a))
and $\Delta_{sc}$ (\figref{fig3}(b)). (Note that $F^\text{crit}$
denotes \emph{all} critical points, including local and global maxima.)
Here, $\Delta_{0c}$ is varied by
varying $\omega_0$ while keeping all other frequencies fixed, and
$\Delta_{sc}$ is varied by varying $\omega_{sc}$.
We find that generally there exist two regimes of operation:
a regime where there are two critical efficiencies, 
one at $F^\text{crit}=0.87$ (light red dashed line) and the other at $F^\text{crit}<0.87$ (dark blue dashed line),
and a regime where only the smaller efficiency survives, $F^\text{crit}<0.87$ (dark blue dashed line). 
These two regimes clearly reflect to what extent the detunings can be compensated for by incident pump powers.
In the first regime, the detunings can be completely compensated and the maximal efficiency is 
only limited by cooperativity $C$. The global maximum is explicitly computed from \eqref{dcrit} as $F^\text{crit}=0.87$ and 
can be realized by applying either of the two critical powers (light red solid lines) corresponding to either vanishing or non-vanishing $\delta$'s.
The required power then grows with the increasing $\Delta$'s while the global maximal efficiency stays the same.
In the second regime, \eqref{dcrit} breaks down
(leading to complex solutions), and the critical parameters can only be computed from 
setting ${\partial F \over \partial P_{b1}} =
0,~{\partial F \over \partial P_{b2}} = 0$. 
Clearly, in this regime, the detunings cannot be fully compensated so that the maximal efficiency falls off rapidly.

\section{Difference frequency generation}
\label{secDFG}

\begin{figure}[h!]
\centering
\includegraphics[scale=0.6]{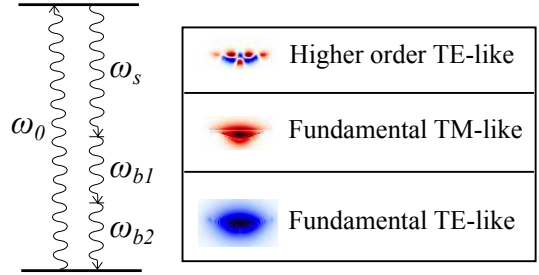}
\caption{Left panel: a diagrammatic representation of difference frequency generation process (DFG), 
Right panel: $E_r$ and $E_z$ components of a higher-order
TE-like mode (for the emitter photon),
 a fundamental TM0-like mode (for the signal photon) and
a fundamental TE0-like mode (for the pump photons).  
TE and TM characters of the modes are ill-defined since the
mirror symmetry in $z$ (the out-of-plane direction perpendicular
to the plane of the resonator) is strongly broken.
Each mode possesses appreciable in-plane and out-of-plane electric
field components, which lead to a non-vanishing $\beta$ (see \eqref{beta}).
\label{fig:fig5}}
\end{figure}

So far, we have tailored our resonator design to the FWM-BS process. Designing a cavity for the alternative scheme of DFG is significantly
more challenging since the pump modes must be designed at a wavelength of $\sim 3\mathrm{\mu m}$, more than thrice that of the SiV emitter, 
so that the generated signal falls within the telecom band (see \figref{fig5}).
For these wavelengths, phase-matching can only be achieved by utilizing higher-order waveguide modes; as a consequence, the overlap between the cavity modes suffers, 
degrading the $\beta$ coefficient. Nevertheless, with the help of the phase-matching method described in Appendix~\ref{design}, we were able to
 obtain realistic designs that still yield high efficiencies at reasonable pump powers. Such a design is shown in \figref{fig5}.

In order to separate TE-like and TM-like bands as wide as possible, we choose a shallow etch angle of $\theta=70^\circ$; the radius and height are
$34~\mathrm{\mu m}$ and $463~\mathrm{nm}$ respectively. The waveguide dimensions are chosen to be larger than the previous design for FWM-BS 
in order to host high-Q modes at very different wavelengths: 
$\left(m_0=461,\lambda_{0c} = 0.738~\mathrm{\mu m}\right)$, $\left(m_s=235,\lambda_s=1.567~\mathrm{\mu m} \right)$ and
$\left(m_{b1}=m_{b2}=m_b=113,\lambda_{b1c}=\lambda_{b2c}=\lambda_b=2.789~\mathrm{\mu m}\right)$. (Note the pumps are degenerate so that we are actually dealing with
three cavity modes instead of four.) Here, it should be noted that even though we make use of 
two TE-like modes and one TM-like modes (see \figref{fig5}), we found that the overlap $\beta$ does not vanish because 
TE and TM characters of the modes are ill-defined for the triangular waveguide, all the modes having siginficant in-plane and out-of-plane components.
From FDTD, radiation Q's are found to be in excess of $10^8$; we assume operational Q's of $10^6$.
The mode volume of the 0th cavity mode is $V_{0c} \approx 1052 ~{\left(\lambda \over n\right)}^3$, yielding a cooperativity factor of about 50 (assuming
100\% internal quantum efficiency of the emitter). The analysis of DFG exactly follows that of DFWM, except that we interchange $\omega_{b1}\rightarrow -\omega_{b1}$
and $\vec{E}_{b1} \rightarrow \vec{E}^*_{b1}$. While the self- and cross-phase modulation coefficients are comparable to those in the case 
of FWM-BS, the overlap coefficient is about $2.5\times10^{-4}~\mathrm{(\mu J)^{-1}}$, which is more than two orders of 
magnitude smaller than in the case of FWM-BS. With these numbers in place, we calculated the critical efficiency and powers which are found to be $64\%$ and $\sim 1~\mathrm{W}$.  

\section{Conclusion}
We have presented efficiency and power requirements for frequency down-conversion of SiV single photons to telecom wavelengths. Our cavity designs are tailored for a potential all-diamond monolithic approach which directly utilizes the nonlinear optical properties of diamond. This stands in contrast to more common hybrid approaches~\cite{McCutcheon09} which employs an external nonlinear material, usually a $\chi^{(2)}$ medium, for frequency conversion. Although the power requirements (on the order of $10-100~\mathrm{mW}$) are greater for all-diamond systems since the inherent \ch nonlinearity of diamond is weaker, the maximum-achievable efficiencies are comparable to $\chi^{(2)}$-based systems, for example Ref.~\cite{McCutcheon09}. The advantage of a potential all-diamond approach is the superior quality of the native emitters in monolithic diamond as well as relative ease of fabrication, scalability and high throughput promised by the advent of angle-etched techniques \cite{Burek12}.

\begin{appendices} 


\section{Nonlinear terms in Hamiltonian}
\label{hamiltonian}
The perturbative nonlinear interaction energy is given by \cite{Hillery09}
\begin{align}
\delta U^\text{NL} = - {1 \over 2} \int~dV~\vE^*(\omega) \cdot \mathbf{P}^\text{NL}(\omega). \label{eq:UNL}
\end{align}
There are detailed theoretical methods on how to rigorously quantize a
nonlinear system \cite{Drummond90,Hillery09} but we will not follow
those arguments here since we are only interested in the perturbative
regime.

The nonlinear polarization for four-wave mixing Bragg scattering
process is given in \cite{Boyd92},
\begin{align}
  P_i^\text{NL}(\omega_s) = D \epsilon_0 \chi^{(3)}_{ijkl}(\omega_s;
  \omega_0, \omega_{b1}, -\omega_{b2})
  E_{0j}E_{b1k}E_{b2l}^*, \label{eq:PNL}
\end{align}
where $D$ is the permutation factor and in this case $D=6$
\cite{Boyd92}. The electric fields shoulqd be normalized as
\begin{align}
  \int~dV~\vE_{\mu}^*(\omega) \cdot \epsilon \vE_{\mu}(\omega) &= {1 \over 2} \hbar \omega_{\mu} \\
  \int~dV~\vE_{b}^*(\omega) \cdot \epsilon \vE_{b}(\omega) &= {1 \over
    2} |a_b|^2
\end{align}
In an isotropic medium like diamond, the components of the nonlinear
susceptibility tensor are given by $\chi_{ijkl} = \chi_{xxyy}
\delta_{ij} \delta_{kl} + \chi_{xyxy} \delta_{ik} \delta_{jl} +
\chi_{xyyx} \delta_{il} \delta_{jk}$. Substituting \eqref{PNL} in
\eqref{UNL} leads to the expression for $\beta$ as given in
\eqref{beta}. Similar calculations can be performed to yield the
expressions for $\alpha$'s, \eqreftwo{aub}{ab1b2}. For wavelengths of
interest which are well into non-resonant regime,
 full-permutation symmetry holds \cite{Boyd92} and 
we approximate that $ \chi_{xxyy} \approx \chi_{xyxy} \approx
\chi_{xyyx} \approx \chi_{xxxx}/3 = \ch/3$.
For diamond, $\ch = 2.5\times 10^{-21}~\mathrm{m^2/V^2}$
\cite{Boyd92}.

\section{Adiabatic elimination}
\label{adiabatic-elimination}

When $\Omega$ is sufficiently small, $c_r$ becomes the slowest
dynamical variable in the system leading the other three. In this
case, $c_e,~c_0,~c_s,~c_l$ all decay approximately at the same rate as
$c_r$. In other words, $c_e,~c_0,~c_s,~c_l$ are simply proportional to
$c_r$ (they follow $c_r$ except for a constant pre-factor). To obtain these
proportionalities, we can formally set
$\dot{c}_e,~\dot{c}_0,~\dot{c}_s,~\dot{c}_l = 0$ individually (adiabatic
elimination) and solve for these variables as functions of
$c_r$. This leads to a simple rate equation for $c_r$:
\begin{align}
\dot{c}_r = ( i \delta_r - { \kappa_r \over 2} ) c_r,
\end{align}
where $\delta_r$ is some phase factor giving rise to unitary
oscillations and $\kappa_r$ is the effective rate of population loss
predicted from the state $\ket{r}$. The latter is given by \small
\begin{align}
  \kappa_r = {|\Omega|^2 \over \gamma} \left( 1 - {4 |\beta|^2 C
      \kappa_0 \kappa_s \over 16 |\beta|^2 + 8 |\beta|^2 \left(
        \kappa_0 \kappa_s \left( 1 + C \right) - 4 \delta_{0c}
        \delta_{sc} \right) + \left( 4 \delta_{0c}^2 + \left( 1 + C
        \right)^2 \kappa_0^2 \right) \left(4 \delta_{sc}^2 +
        \kappa_s^2 \right) }\right).
\end{align}
\normalsize Self-consistency of adiabatic elimination requires that
the population loss out of $r$ does not exceed the rate $\kappa_s$
with which a photon can leak out through the cavity mode $s$,
$\kappa_r < \kappa_s$. Additionally, in our calculations, we have
rigorously checked the validity of adiabatic elimination by direct
comparison with exact numerical integration of \eqreftwo{rate1}{ds},
using $|\Omega| \sim 0.1 \kappa_0$.

\section{Ring resonator design}
\label{design}

\begin{figure}[h!]
\centering
\includegraphics[scale=0.5]{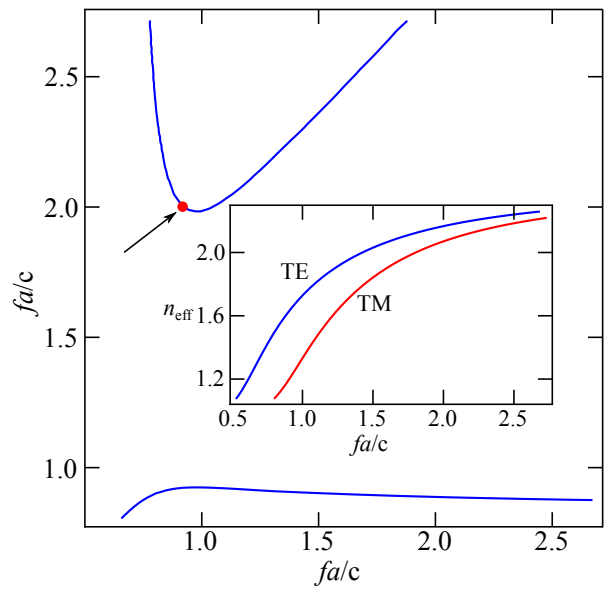}
\caption{Phase-matching diagram for the FWM-BS process in a triangular waveguide 
of $h=0.25a$ and $\theta=60^\circ$. The red circle indicates the frequencies used in the design 
of ring resonator. Inset: fundamental TE0 and TM0 bands 
computed by MPB. $a$ is an arbitrary normalization length to be chosen later (see text). \label{fig:fig6}}
\end{figure}

We begin by examining the modes of an infinite straight waveguide having a
triangular cross-section. Specifically, 
we chose an etch-angle of $60^\circ$ and a height of $0.25 a$
where $a$ is an arbitrary normalization length to be chosen later.
The modes of the waveguides can be quickly computed 
by standard band structure solvers;
for our simulations, we use the freely-available open-source MPB \cite{Johnson2001:mpb} 
which compute eigenfrequencies $f$ (in units of $c/a$) 
at specified $k$'s (in units of $2 \pi /a$).
We select two bands which are
far-separated in frequency but still have significant overlap; such bands are 
readily afforded by the fundamental TE0 and fundamental TM0 modes (see the inset of \figref{fig6}).
Next, we define the phase mismatch 
as a function of frequencies (for the case of FWM-BS)
\begin{align}
  \delta k(f_1,f_2,f_3,f_4) = k_1 + k_2 - k_3 - k_4.
\end{align}
We want two of the frequencies, say, $f_2$ and $f_4$ to lie 
within the NIR band and the other two, $f_1$ and $f_3$, within the telecom band.
Therefore, we identify: $k_1 = k_\text{TE0}(f_1),~k_2 = k_\text{TM0}(f_2),~
k_3 = k_\text{TE0}(f_3),~k_4 = k_\text{TM0}(f_4)$.

$\delta k$ is a function of four independent variables $f_1,~f_2,~f_3,~f_4$
which is hard to visualize. We subject the frequencies to 
the constraint $f_1 + f_2 = f_3 + f_4$.
 In order to reduce the number of free variables further, we make
$f_3$ a function of $f_1$ and $f_2$, i.e., $f_3 =
f_3(f_1,f_2)$. This also makes $f_4$ a function of $f_1$ and
$f_2$. The functional form of $f_3$ can be chosen arbitrarily. A
simple intuitive choice would be to set $f_3 = f_1 - \delta f$ for some
offset $\delta f$. Since the number of variables has been reduced to two, we
are ready to visualize $\delta k$. Choosing $df = 0.1$, 
 we plotted a contour of $\delta k = 0$ over
the $f_1$ vs. $f_2$ plane (see \figref{fig6}). Each point along the
contour yields four waveguide modes with frequencies that are
perfectly phase-matched for an efficient FWM-BS process. 
In particular, we chose a point on the contour
(red circle in \figref{fig6}): $f_1=0.95,~f_2=2.00,~f_3=0.85,~f_4=2.10$.
It is easily noticed that $f_4$ and $f_1$ have the approximate 
ratio as two frequencies from SiV and telecom bands. So we identify
$f_4 {c \over a} = f_0 = { c \over 0.738~\mathrm{\mu m} }$, yielding $a=1.55~\mathrm{\mu m}$ and
$f_s = f_1 {c \over a} = {c \over 1.63~ \mathrm{\mu m} }$.
This also specifies the height of the waveguide, $h \approx 390~\mathrm{nm}$.
Remembering that ring
resonances satisfy the condition $m_i = 2 \pi R n_{\text{eff},i}
f_i,~i \in \{0,s,b1,b2\}$, we can find an optimal $R$ which makes
$m_i$ an integer (or closest to an integer) for each of the four
modes. We found that $R \sim 14.7\mathrm{\mu m}$ approximately gives
$m_0 = 262,~m_s=95,~m_{b1}=79,~m_{b2}=246$, also satisfying the
condition $m_0 + m_{b1} = m_s + m_{b2}$. Now we can 
proceed to simulate the ring resonator via full-scale FDTD, 
using the predicted design parameters.
A quick inexpensive search (on the order of ten iterations)
in the vicinity of the above parameters directly yields the design 
given in Section~\ref{secDesign}.

A similar procedure is followed to design a cavity for DFG except that the angle is made shallower $\theta=70^\circ$.
We found that the 8th order TE-like band together with fundamental TE-like and TM-like bands yield 
a regime of phase-matched frequencies (not shown) similar to the one depicted in \figref{fig6}, thereby identifying 
appropriate parameters which lead to the resonator design of Section~\ref{secDFG} after a quick numerical search.

\end{appendices}
\end{document}